\documentclass{emulateapj}
\usepackage{apjfonts}

\usepackage{amssymb}
\usepackage{natbib}

\usepackage{xspace}
\usepackage{graphicx}

\newcommand{\err}[2]{\ensuremath{^{+#1}_{-#2}}\xspace}

\slugcomment{Received 2005 June 24; accepted 2005 October 10}

\shorttitle{Cyclotron lines in V\,0332$+$53}
\shortauthors{Pottschmidt et al.}

\begin{document}

\title{\textsl{RXTE} Discovery of Multiple Cyclotron Lines during the 2004
  December Outburst of V\,0332$+$53}

\author{Katja~Pottschmidt\altaffilmark{1},
Ingo~Kreykenbohm\altaffilmark{2,3},
J\"orn~Wilms\altaffilmark{4},
Wayne~Coburn\altaffilmark{5},
Richard~E.~Rothschild\altaffilmark{1},
Peter~Kretschmar\altaffilmark{6},
Vanessa~McBride\altaffilmark{7},
Slawomir~Suchy\altaffilmark{1},
R\"udiger~Staubert\altaffilmark{2}} 

\altaffiltext{1}{University of California, San Diego, Center for
  Astrophysics and Space Sciences, 9500 Gilman Dr., La Jolla, CA
  92093-0424, USA, \{kpottschmidt,rrothschild,ssuchy\}@ucsd.edu}
  \altaffiltext{2}{Institut f\"ur Astronomie und Astrophysik,
  Astronomy Section, Sand 1, 72076 T\"ubingen, Germany,
  staubert@astro.uni-tuebingen.de} \altaffiltext{3}{\textsl{INTEGRAL}
  Science Data Centre, Chemin d'\'Ecogia 16, 1290 Versoix,
  Switzerland, ingo.kreykenbohm@obs.unige.ch}
  \altaffiltext{4}{Department of Physics, University of Warwick,
  Coventry CV4 7AL, United Kingdom, j.wilms@warwick.ac.uk}
  \altaffiltext{5}{Space Sciences Laboratory, University of
  California, Berkeley, Berkeley, CA 94702-7450, USA,
  wcoburn@ssl.berkeley.edu} \altaffiltext{6}{European Space Agency,
  European Space Astronomy Centre, Villafranca del Castillo, 28080
  Madrid, Spain, peter.kretschmar@esa.int} \altaffiltext{7}{School of
  Physics and Astronomy, Southampton University, Southampton SO17 1BJ,
  United Kingdom, vanessa@astro.soton.ac.uk}

\email{kpottschmidt@ucsd.edu}

\begin{abstract}
  We present an analysis of the 2--150\,keV spectrum of the transient
  X-ray pulsar V\,0332$+$53 taken with the Rossi X-Ray Timing Explorer
  (\textsl{RXTE}) in 2004 December. We report on the detection of
  three cyclotron resonance features at 27, 51, and 74\,keV in the
  phase-averaged data, corresponding to a polar magnetic field of
  $2.7\times 10^{12}$\,G. After 4U\,0115+63, this makes V\,0332+53 the
  second accreting neutron star in which more than two cyclotron lines
  have been detected; this has now also been confirmed by
  \textsl{INTEGRAL}.  Pulse-phase spectroscopy reveals remarkably
  little variability of the cyclotron line through the 4.4\,s X-ray
  pulse.
\end{abstract}


\section{Introduction}

The recurring transient X-ray pulsar V\,0332+53 was discovered in 1983
in \textsl{Tenma} observations made from 1983 November to 1984 January
\citep{tanaka83a,makishima90a}. Subsequently, an outburst was revealed
in \textsl{Vela 5B} data to have occurred in 1973 summer
\citep{terrell83a,terrell84a}. In rapid follow-up observations to the
\textsl{Tenma} detection by \textsl{EXOSAT} during the 1983 outburst,
4.4\,s pulsations were discovered, an accurate position was
determined, and indications were found for an orbital period of
34.25\,days \citep{stella85a}. The position and subsequent
optical/X-ray variability secured the identification of the
counterpart with the heavily reddened B star BQ~Cam \citep{Argyle83a}.
Later investigations refined the spectral type to O8--9
\citep{negueruela99a}.

Analysis of the \textsl{Tenma} data revealed a shape similar to that
seen in other accreting X-ray pulsars with a flat power law and
exponential cutoff, and also showed evidence of a cyclotron line at
$\sim$28\,keV \citep{makishima90a}. In 1989 September the source
experienced another outburst, this time observed by \textsl{Ginga}
\citep{makino89a}. With the energy range of the Large Area Counters
adjusted to cover the 2--60\,keV range, cyclotron resonant scattering
features (CRSF) were detected at 28.5 and 53\,keV
\citep{makishima90b}. At the time, this was the fourth accreting X-ray
pulsar system to exhibit cyclotron features. Since then a significant
number of this class of objects have been shown to have CRSFs, with
4U~0115$+$63 setting the record for the number of harmonics at five
\citep[see][for a review]{heindl04a}.

V\,0332+53 went into outburst again in 2004 November, reaching a
1.5--12\,keV intensity of $\sim$1\,Crab \citep{remillard04a}. A long
series of observations by the Proportional Counter Array
\citep[PCA;][]{jahoda:05a}\footnote{See also
  \url{http://lheawww.gsfc.nasa.gov/docs/xray/xte/pca/}.} and High
Energy X-Ray Timing Experiment \citep[HEXTE;][]{rothschild:98a} aboard
\textsl{RXTE} were made throughout the outburst. Following up on our
initial announcement \citep{coburn05a}, here we report the discovery
of a third cyclotron line at $\sim$74\,keV, along with clear evidence
for a non-Gaussian shape of the fundamental. In \S2 we present the
observations and the data extraction. The phase-averaged and the
phase-resolved spectra are analyzed and discussed in \S3 and \S4,
respectively. In \S5 we summarize the results and present our
conclusions.

\section{Observations and Data Extraction}

\begin{figure}
\plotone{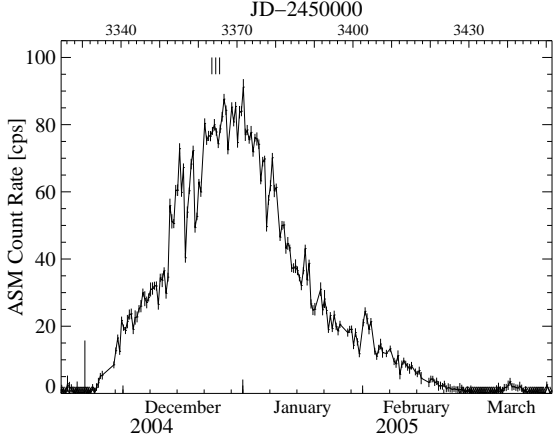}
\caption{\textsl{RXTE} ASM 1.5--12\,keV light curve of the 2004/2005
  outburst of V\,0332+53, binned to 0.5\,day resolution. Tick marks
  denote the times of the observations analyzed in this Letter.}
\label{fig:asm}
\end{figure}

Since the luminosity of the source is strongly variable during the
outburst, we decided to concentrate on the three longest observations
during the peak of the outburst as observed in the soft X-ray band by
the \textsl{RXTE}-All Sky Monitor (ASM; Fig.~\ref{fig:asm}). Results
from the full monitoring data set will be presented elsewhere. We
combined the PCA and HEXTE data taken 2004 December 24--26
(\textsl{RXTE} ObsIDs 90089-11-05-08G, 90089-22-01-00G, and
90089-22-01-01G), giving a total of $\sim$58\,ks of exposure in the
PCA and 14.4\,ks of dead-time-corrected live time in HEXTE cluster B.
At the time of the selected observations, HEXTE's cluster A was not
rocking due to a single event upset, and we discarded data from this
cluster\footnote{Cluster~A was subsequently rebooted and continues to
work normally.}. The data were reduced with HEASOFT version 5.3.1, and
spectral modeling was performed with XSPEC version 11.3.1w
\citep{arnaud:96a}.  For modeling the phase-averaged spectra
(Fig.~\ref{fig:cycl}a), PCA data from 3--20\,keV and HEXTE data from
18--150\,keV were used, and the HEXTE data were rebinned by factors of
3 and 5 between 70--100 and 100--150\,keV, respectively.  A systematic
error of 0.5\% was assumed for the PCA.

\section{Phase-averaged Spectra}

The X-ray spectrum observed from accreting X-ray pulsars is the
superposition of X-rays originating in the accretion column and the
hot spots on the surface of the neutron star. The neutron star's
X-rays are partially intercepted by the material present in the
system, such as the stellar wind, giving rise to significant
absorption which can exceed $10^{23}\,\rm{H-atoms}\,\rm{cm}^{-2}$
(see, e.g., \citealt{becker05a} for a discussion of the formation of
the X-ray continuum in the accretion column).

The ${\cal O}(10^{12}\,\rm{G})$ magnetic field close to the magnetic
poles of the neutron star leads to the formation of CRSFs in the X-ray
spectrum.  The observer's frame energy of the CRSFs is given by
\citep{meszaros92a}

\begin{eqnarray}
\label{eq:cyce}
E_{\rm{C},n} & = \frac{2nE_{\rm F}}{1+z} \left(1 + \sqrt{1 + 2n \frac{E_{\rm F}}{m_{\rm{e}} c^2}\sin^2\theta} \right)^{-1} \\  
\label{eq:nonrel} & \sim \frac{nE_{\rm F}}{1+z},
\end{eqnarray}
where $n=1$,2,\ldots, is the harmonic number, $E_{\rm F} = 11.6\,{\rm
keV} \cdot (B/10^{12}\,{\rm G})$ is the fundamental nonrelativistic
cyclotron energy ($n=1$), $\theta$ is the angle between the wave
vector of the incoming photon and the magnetic field $B$, and
$z\sim0.3$ is the gravitational redshift (assuming a $1.4\,M_\odot$
neutron star with a 10\,km radius). The great importance of CRSF
observations is that they provide the only direct way to measure
neutron star $B$-field strengths.

The lack of adequate theoretical continuum models for accreting
neutron stars necessitates the use of empirical models to describe the
observations \citep{kreykenbohm99a}.  We use a power law modified at
higher energies by a ``Ferm- Dirac cutoff'' \citep{tanaka86a} to fit
the PCA and HEXTE data simultaneously, including a multiplicative
constant as a fitting parameter to allow for the slight difference
($\lesssim$5\%) in the flux calibration of both
instruments. Photoelectric absorption and a strong Fe K$\alpha$
fluorescence line at 6.35\,keV from neutral iron, with a width of
$\sigma_{\rm{Fe K\alpha}}=0.43$\,keV and an equivalent width of
123\,eV are also taken into account. The soft continuum parameters
found (see below) are in general agreement with earlier results
\citep[e.g.,][]{unger92a}.

\begin{figure}
\plotone{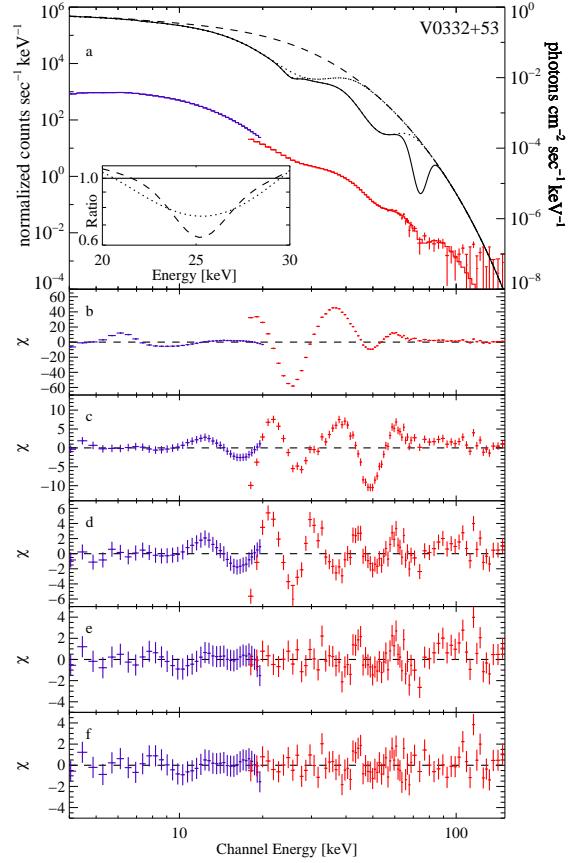}
\caption{Cyclotron line modeling of the phase averaged PCA/HEXTE
  spectrum of V\,0332+53. (a) Combined PCA/HEXTE spectrum (crosses),
  best fit model (histogram) and unfolded spectrum (dotted and solid
  lines, right $y$-axis label), illustrating the cumulative effect of
  the CRSFs on the continuum (see text for a description of the
  continuum model). Note that the lines are already visible in the raw
  count spectrum. The inset shows the best-fit spectra with one
  (dotted line) and two (dashed line) line components for the
  fundamental line relative to the best-fit continuum with no lines
  (solid line).  Panels (b)--(f) show residuals for models taking
  different numbers of CRSFs into account (see Table~1 for the best
  fit values). (b) No line; (c) one line at 25.5\,keV; (d) two lines,
  one at 26.6\,keV and one at 50\,keV; (e) two lines, a fundamental
  line that is best modeled with two Gaussian components at 25.2 and
  26.6\,keV and its harmonic at 49.9\,keV; (f) three lines, the
  fundamental modeled by two components at $\sim$25 and $\sim$27\,keV,
  and two lines at $\sim$51 and $\sim$74\,keV. [\textsl{See the
  elctronic edition of the Journal for a color version of this
  figure.}]}\label{fig:cycl}
\end{figure}

While this model fits the overall shape of the continuum reasonably
well, strong absorption-like features are present above 20\,keV
(Fig.~\ref{fig:cycl}b), which we model as lines with Gaussian optical
depth profiles of the form
\begin{equation}\label{gabs}
\tau(E) = \tau_{\rm{C}}
   \cdot  \exp\left(
 -\frac{1}{2}\left(\frac{E-E_{\rm{C}}}{\sigma_{\rm{C}}}\right)^2
                \right),
\end{equation}
where $E_{\rm{C}}$ is the line energy, $\sigma_{\rm{C}}$ is the line
width, and $\tau_{\rm{C}}$ is the optical depth at the line
center. The spectral shape is then given by multiplying the continuum
spectrum by $\exp(-\tau(E))$ \citep[][and references
therein]{kreykenbohm04a}.

We start by modeling the strongest feature between 20 and 30\,keV with
one of these Gaussian components (see Table~1 for the relevant best
fit parameters). This feature corresponds to the CRSF first reported
by \citet{makishima90b}. While the 20--30\,keV residuals improve
significantly compared to the pure continuum fit when this component
is included, another strong and remarkably well defined
absorption-like feature emerges at $\sim$50\,keV
(Fig.~\ref{fig:cycl}c).  This feature is the first harmonic of the
cyclotron line, found at $\sim 2$ times the energy of the fundamental
and already tentatively reported by \citet{makishima90b}. Taking these
two lines into account, weaker residuals remain at $\sim$25\,keV
(Fig.~\ref{fig:cycl}d). Similar to 4U\,0115+63 \citep{heindl99a} and
other sources, these residuals are caused by the fundamental CRSF
having a non-trivial shape, which cannot be modeled well with any
simple line shape (see \S\ref{sec:summary}). To describe the complex
shape we add a shallow ($\tau_{\rm{cyc,1b}}\sim0.3$) and narrow
($\sigma_{\rm{cyc,1b}}\sim 1.5$\,keV) line component at an energy
similar to that of the fundamental,
$E_{\rm{cyc,1b}}=25.2\err{0.2}{0.2}$\,keV; the shape of this more
complex line profile is shown in the inset of Fig.~\ref{fig:cycl}a.
Although lowering the $\chi^2$ considerably, the fit is still not
acceptable, and significant residuals remain (Fig.~\ref{fig:cycl}e).
Adding another line at $73.7\err{1.5}{1.3}$\,keV improves the fit
significantly to $\chi^2_{\rm{red}} = 1.04$ (Fig.~\ref{fig:cycl}f, the
unfolded spectrum and the consecutive effect of the individual CRSFs
are shown in Fig.~\ref{fig:cycl}a).  The $F$-test probability that
this improvement is due to chance is only $3\times 10^{-6}$ (however,
see \citealt{protassov02a} concerning the applicability of the
$F$-test in this case).

\begin{deluxetable}{lrrrr}
\tablecaption{Spectral parameters for the phase-averaged
spectrum. \label{tab:values}} \tablecolumns{5} \tablehead{Parameter &
2c & 2d & 2e & 2f} \startdata $N_{\rm{H}}$ ($10^{22}\,\rm{cm}^{-2}$) &
$0.0^{+0.2}_{-0.0}$ & $1.3^{+0.5}_{-0.3}$ & $1.5^{+0.5}_{-0.7}$ &
$1.6^{+0.6}_{-0.6}$ \\[1mm] $\Gamma$ & $-0.16^{+0.02}_{-0.01}$ &
$0.04^{+0.04}_{-0.03}$ & $0.37^{+0.07}_{-0.07}$ &
$0.42^{+0.05}_{-0.05}$ \\[1mm] $E_{\rm{cut}}$ (keV) &
$0.0^{+0.1}_{-0.0}$ & $0.0^{+0.7}_{-0.0}$ & $13.0^{+5.9}_{-4.2}$ &
$17.2^{+3.2}_{-4.6}$ \\[1mm] $E_{\rm{fold}}$ (keV) &
$6.1^{+0.0}_{-0.0}$ & $7.1^{+0.2}_{-0.1}$ & $7.0^{+0.2}_{-0.2}$ &
$8.0^{+0.7}_{-0.5}$ \\[1mm] $E_{\rm{cyc,1a}}$ (keV) &
$25.5^{+0.0}_{-0.0}$ & $26.3^{+0.1}_{-0.1}$ & $26.6^{+0.1}_{-0.1}$ &
$27.1^{+0.2}_{-0.1}$ \\[1mm] $\sigma_{\rm{cyc,1a}}$ (keV) &
$4.7^{+0.0}_{-0.0}$ & $5.5^{+0.1}_{-0.1}$ & $7.0^{+0.5}_{-0.3}$ &
$7.6^{+0.2}_{-0.2}$ \\[1mm] $\tau_{\rm{cyc,1a}}$ &
$1.02^{+0.01}_{-0.01}$ & $1.22^{+0.03}_{-0.04}$ &
$1.38^{+0.20}_{-0.11}$ & $1.81^{+0.08}_{-0.06}$ \\[1mm]
$E_{\rm{cyc,1b}}$ (keV) & \nodata & \nodata & $25.2^{+0.2}_{-0.2}$ &
$25.2^{+0.2}_{-0.2}$ \\[1mm] $\sigma_{\rm{cyc,1b}}$ (keV) & \nodata &
\nodata & $1.3^{+0.5}_{-0.5}$ & $1.5^{+0.3}_{-0.5}$ \\[1mm]
$\tau_{\rm{cyc,1b}}$ & \nodata & \nodata & $0.35^{+0.13}_{-0.04}$ &
$0.34^{+0.09}_{-0.03}$ \\[1mm] $E_{\rm{cyc,2}}$ (keV) & \nodata &
$50.0^{+0.4}_{-0.4}$ & $49.9^{+0.4}_{-0.3}$ & $51.3^{+0.4}_{-0.4}$
\\[1mm] $\sigma_{\rm{cyc,2}}$ (keV) & \nodata & $6.5^{+0.5}_{-0.3}$ &
$6.5^{+0.5}_{-0.5}$ & $8.9^{+0.6}_{-0.6}$ \\[1mm] $\tau_{\rm{cyc,2}}$
& \nodata & $0.94^{+0.09}_{-0.09}$ & $1.19^{+0.14}_{-0.11}$ &
$2.21^{+0.31}_{-0.26}$ \\[1mm] $E_{\rm{cyc,3}}$ (keV) & \nodata &
\nodata & \nodata & $73.7^{+1.5}_{-1.3}$ \\[1mm] $\sigma_{\rm{cyc,3}}$
(keV) & \nodata & \nodata & \nodata & $4.5^{+2.5}_{-2.7}$ \\[1mm]
$\tau_{\rm{cyc,3}}$ & \nodata & \nodata & \nodata &
$3.3^{+2.6}_{-1.1}$ \\[1mm] $\chi^2_{\rm{red}}/\rm{dof}$ & $16.24/104$
& $4.05/101$ & $1.37/98$ & $1.04/95$ \enddata \tablecomments{ The
columns refer to residuals shown in Fig.~2. Symbols denote the
equivalent hydrogen column, $N_{\rm{H}}$, the photon index, $\Gamma$,
the cutoff and folding energies of the Fermi-Dirac continuum,
$E_{\rm{cut}}$ and $E_{\rm{fold}}$, and the cyclotron line parameters
$E_{\rm{cyc}}$, $\sigma_{\rm{cyc}}$, and $\tau_{\rm{cyc}}$ (defined in
Eq.~\ref{gabs}) for fits with 1, 2, and 3 CRSFs. CRSF parameters
labeled 1a and 1b correspond to the complex profile of the fundamental
line. See text for a description of the continuum
model. $\chi^2_{\rm{red}}$ is the reduced $\chi^2$ and dof is the
number of degrees of freedom. All uncertainties are 90\% confidence
levels for one interesting parameter ($\Delta\chi^2=2.71$).  }
\end{deluxetable}

To verify the detection of the CRSFs and to confirm their independence
from the choice of the continuum, we modeled the data with other
continua commonly used to describe the spectra of accreting pulsars.
The negative positive exponential continuum
\citep[NPEX,][]{mihara95a}, for example, also requires the presence of
three CRSFs at 27, 51.7, and 74.8\,keV, again with a non-Gaussian
fundamental. Other continua gave similar results, as do other
prescriptions for the shape of the CRSF.  The actual choice of the
continuum or line shape is thus irrelevant for the detection and
characterization of the CRSFs at the level of the \textsl{RXTE} energy
resolution. We therefore report the discovery of the second harmonic
of the cyclotron line in V\,0332+53 with \textsl{RXTE}, confirming our
preliminary initial report for this data set \citep{coburn05a}. This
discovery has been independently confirmed by subsequent
\textsl{INTEGRAL} observations \citep{kreykenbohm05a}.

\section{Phase-resolved Spectra}

\begin{figure}
\plotone{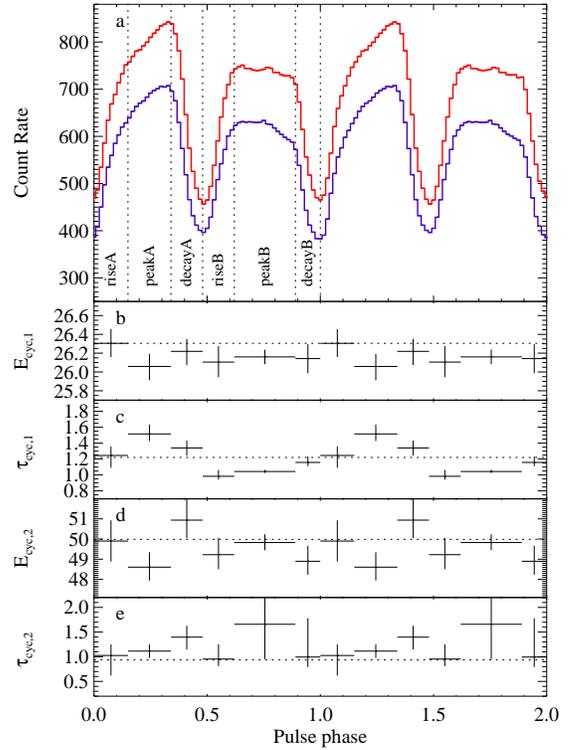}
\caption{Pulse profile and CRSF variability of V\,0332+53. (a) HEXTE
  (full band, upper profile) and PCA (20--50\,keV; lower profile)
  pulse profile. Also shown are the phase evolution of (b) the energy
  and (c) the depth of the fundamental and (d, e) the first harmonic
  cyclotron line. Error bars are at the 90\% level for one interesting
  parameter; dotted lines indicate the best-fit values from the two
  CRSF fit to the phase averaged spectrum. [\textsl{See the elctronic
  edition of the Journal for a color version of this figure.}]}
\label{fig:profile}
\end{figure}

The neutron star's magnetic axis is offset from its spin axis, not
only giving rise to X-ray pulsations but also confronting the observer
with a complex situation. Due to the rotation of the neutron star, the
accretion column and the hot spots are seen under a constantly
changing viewing angle. Since the physical conditions are likely to be
variable within the emission region and since the CRSF line shape
depends on the viewing angle, we expect the X-ray continuum and line
parameters to depend on the pulse phase
\citep{araya00a,araya99a,isenberg98a}. This has been confirmed by
pulse-phase-resolved spectroscopy in a number of sources, such as
GX~301$-$2 \citep[][finding $\sim$20\% change in the line
energy]{kreykenbohm04a}, Vela~X-1 \citep{kreykenbohm02a,barbera03a},
or 4U~0115+63 \citep{mihara04a}.

To obtain phase-resolved spectra, we determined the pulse period for
each of the three observations using standard techniques. This
approach is necessary, as the accumulated uncertainty of the only
available orbital ephemeris of the neutron star \citep{stella85a} is
too large to allow any meaningful orbit correction.  The average
non-orbit-corrected X-ray period is $P=4.3745$\,s. After
phase-aligning the resulting pulse profiles, we determined
phase-resolved spectra.  Fig.~\ref{fig:profile}a shows the resulting
HEXTE and PCA pulse profile of V\,0332+53.

Since the data modes of the observation do not contain a PCA data mode
with energy resolution below 20\,keV and a sufficiently high time
resolution for phase-resolved spectroscopy, we performed phase
resolved spectroscopy using the HEXTE data only. Signal-to-noise ratio
considerations allow a maximum of six phase bins
(Fig.~\ref{fig:profile}a). In the resulting phase-resolved spectra,
the source is detected to $\sim$100\,keV, such that we only model the
lower two CRSFs and do not include the structure of the fundamental
line.  The absence of soft spectral data forces us to hold $\Gamma$
and $E_{\rm{cut}}$ at the values found from the appropriate
phase-averaged spectrum. This is justified by the relative constancy
of these parameters as seen with earlier instruments
\citep{unger92a}. Both CRSFs are detected with high significance in
all pulse phases.  Figs.~\ref{fig:profile}b--e show the evolution of
the CRSF line energy and depth over the X-ray pulse.  In contrast to
the sources mentioned above, there is little to no variation in these
parameters, no other spectral parameters show any significant
variation either.

\section{Summary and Conclusions}\label{sec:summary}

We have presented the discovery of a third cyclotron line in
V\,0332+53, making this X-ray pulsar the second, after 4U\,0115+63,
with more than two cyclotron lines, and confirming the identification
of the 26\,keV line as the fundamental CRSF. Similar to sources such
as 4U\,0115+63 or GX~301$-$2, the HEXTE spectrum shows that the
profile of the fundamental line has an asymmetric profile that is
shallower toward lower energies. Such a profile is consistent with
expectations from the asymmetric shape of the relativistic cross
section for resonant electron scattering, but could also be influenced
by the emission wings predicted in Monte Carlo simulations
\citep{araya00a,isenberg98a} or by a variation of the $B$-field in the
emission region.

Consistent with the trend seen in most sources with multiple CRSFs,
the energies of the fundamental and its harmonics do not scale like
1:2:3, as predicted by the non-relativistic Eq.~\ref{eq:nonrel}.
Rather, the line ratios are slightly smaller, in agreement with the
prediction of relativistic quantum mechanics (Eq.~\ref{eq:cyce}).
Using all three observed line energies (taking $E=26.3$\,keV for the
fundamental line) and assuming $z=0.3$, for V\,0332$+$53 a most
probable polar magnetic field of $2.7\times 10^{12}$\,G is found.
Consistency between all three CRSF energies can only be reached for
$\theta\gtrsim 60^\circ$; i.e., photons scattered out of the line have
an initial direction that is perpendicular to the $B$-field, as is
expected from the significant increase of the electron scattering
cross section with $\theta\rightarrow 90^\circ$.  This result provides
a much stronger confirmation of the interpretation of the line
features as CRSFs than is possible in sources where just the
fundamental is observed.

Pulse-phase-resolved spectroscopy revealed that V\,0332+53 shows far
less variability in the CRSF parameters than many other sources.  The
origin of the CRSF variability in these sources is unclear, although
it has been pointed out that their complex pulse profiles make it
likely that different regions of the accretion column or even
different magnetic poles are seen during the X-ray pulse. The
remarkably simple hard X-ray pulse profile of V\,0332+53 suggests a
less complex accretion geometry than in other sources, which might
also explain the very weak CRSF variability.

In conclusion, our observation of multiple lines strongly confirms the
interpretation of these features as cyclotron lines. For a further
interpretation of our results, however, more realistic models for the
neutron star continuum and especially for the line shape are needed,
to include the radiative transfer in the line, multiple photon
scattering, etc., and thus also predict emission wings and strongly
asymmetric profiles.  Work is currently ongoing in our collaboration
to compute such models for a variety of parameters to make such
progress possible in the near future.

\begin{acknowledgments}
  We thank both first referees for their remarks, which greatly
  improved this \textsl{Letter}. We acknowledge the support of NASA
  contract NAS5-30720, NASA grant NAG5-10691, NSF grants INT-9815741
  and INT-0003773, and DLR grants 50OG9601, 50OG0501, and
  50OR0302. V.~McBride is funded jointly by the NRF (South Africa),
  British Council, and Southampton University.
\end{acknowledgments}

\end{document}